\documentclass[aps,prd,preprintnumbers, nofootinbib]{revtex4}
\usepackage{bm}
\usepackage{latexsym}
\usepackage{amsmath,amsfonts,amssymb}
\usepackage{graphicx,epsfig}
\usepackage{psfrag}
\usepackage{amsthm}
\usepackage{color}

\usepackage{amsmath}
\usepackage{braket}
\usepackage{indent first}
\usepackage[normalem]{ulem}
\useunder{\uline}{\ul}{}

\begin{document}
\title{
Information recovery from evaporating black holes
}

\author{Samuel L. Braunstein}
\email{sam.braunstein@york.ac.uk}
\affiliation{Computer Science, University of York, York Y010 5GH, United Kingdom}

\author{Saurya Das}
\email{saurya.das@uleth.ca}
\affiliation{Theoretical Physics Group and Quantum Alberta, Department of Physics and Astronomy, University of Lethbridge, 4401 University Drive, Lethbridge, Alberta T1K 3M4, Canada}

\author{Zhi-Wei Wang}
\email{zhiweiwang.phy@gmail.com}
\affiliation{College of Physics, Jilin University,
Changchun, 130012, People’s Republic of China}

\begin{abstract}
\par\noindent
We show that {the apparent horizon and the region near $r=0$} of an evaporating charged, rotating black hole are timelike. It then follows that  black holes in nature, which invariably have some rotation, 
have a channel, via which classical or quantum information can escape to the outside, while the black hole shrinks in size. 
We discuss implications for the information loss problem.

\end{abstract}

\maketitle

%
\par\noindent
%
Classical and quantum information can enter a black hole's event horizon. 
However, it is normally assumed that what emerges from it at late times is just thermal Hawking radiation carrying minuscule information \cite{hawk2}. 
Therefore, when the black hole evaporates completely, 
all ingoing information is apparently lost forever. 
This in essence is the so-called information loss problem. 
This is demonstrated by the Penrose diagrams in Figures 1 and 2. 
Figure 1 depicts a stationary Schwarzschild (non-rotating, uncharged) black hole.
In this case, the singularity is spacelike, and it is clear that information, from within the horizon, propagating along null (or timelike) geodesics cannot reach the outside Universe. The situation does not improve when this black hole is evaporating, as can be seen from the second diagram in the same figure.
Similarly, Figure 2 shows the 
Penrose diagram of a maximally extended 
rotating Kerr black hole where now the singularity is 
timelike. In this case, although information 
(again propagating along null rays) can exit a future horizon, it does so only to emerge in another Universe. In other words, the information loss problem remains in the present Universe. Two points may be noted here: \\
i. The Penrose diagrams for a rotating black hole, charged black hole and charged {\it and} rotating black hole are virtually identical, and   \\
ii. all black holes in nature (much like other astronomical objects) are rotating and uncharged, and the probability of finding a black hole with zero rotation is practically nil. 
%
{This is supported by theoretical studies \cite{king}}, as well as by the recent gravitational-wave and other observations
\cite{gw,bhobs}. 
%
\footnote{Even if one takes superradiance and Hawking radiation into account, it is highly probable that the black hole at late times evolves to a state of a fixed and non-zero value of $a/M$ (where as usual $a=J/M$ with $J$ and $M$ being the angular momentum and mass of the black hole, respectively), or that it approaches the Schwarchild limit at very late times \cite{page2,hiscock1,hiscock2}. 
}
Therefore, hereafter we will only consider rotating black holes, and
as long as it has some angular momentum, however small, 
the causal structure and our analysis will remain valid for the lifetime of a black hole. 
Furthermore, except near the very end of the lifetime of the black hole, the spacetime curvatures are small and our results are robust and completely trustworthy. 

In particular, in this paper we show that for a black hole which is Hawking radiating, there is a classical channel through which information can escape, and following the above reasoning, it provides an extended window of information recovery from its interior. 
In the process, the black hole shrinks of course, but at a faster rate than 
predicted by Hawking radiation, because of the additional outflow of information and associated matter. 
We will also allow for a non-zero charge, $Q$, 
in our calculations, as this does not give rise to any added complication. 
We demonstrate the above by constructing the Penrose diagram for the above 
process. 
and prove the following: \\
1. 
the region immediately surrounding $r=0$ is timelike, and \\
%
2. The apparent horizon is timelike.\\
Results 1 and 2 imply null geodesics originating 
{from anywhere near the centre of the black hole} 
to the apparent horizon may emerge to the outside Universe. This in turn provides a route for classical or quantum information to escape from the black hole. 
This escape of potentially a large amount of information 
must be taken into account in any attempt to resolve the information loss problem. What is most significant is perhaps the fact that the escaping information is not thermal.  
%
%
%

{We start with the time-dependent rotating and charged Vaidya-type
black hole \cite{Vaidya,jing}. As shown in \cite{pw} and in Appendix \ref{AppendixA}, 
such a metric represents an evaporating black hole, a rotating and charged one in this case:}
\begin{eqnarray}
\!ds^2\!&=&-\biggl(1-\frac{2Mr-Q^2}{\sigma^2}\biggr)du^2 + 2du\,
dr + \sigma^{2}d\theta^2 \label{rotmetric} \nonumber \\
&&- 2a\sin^2\theta\, dr\, d\phi
-\frac{2(2Mr-Q^2) a }{\sigma^2}\sin^2\theta\, du\, d\phi
+\biggl(\frac{2Mr-Q^2}{\sigma^2}\,a^2\sin^2\theta
+r^2 + a^2\biggr)\sin^2\theta\, d\phi^2, 
\end{eqnarray}
where $\sigma^{2}\equiv r^2+a^2\cos^2\theta$, and $M=M(u)$, $Q=Q(u)$
and $a=a(u)$ denote smooth decreasing functions of the retarded time $u$.
%
This ensures that the black hole is indeed evaporating. 
%

In order to draw the two-dimensional Penrose diagram for a rotating black hole, it
is sufficient to restrict the metric to the symmetry axis
along $\theta=0$ \cite{Carter66}. 
This is, from Eq.~(\ref{rotmetric}):
\begin{eqnarray}
ds^2 &=& -\biggl( 1 - \frac{2Mr - Q^2}{r^2 + a^2}\biggr) du^2
{+} 2\, du\, dr \nonumber \\
&=& -\biggl[ \frac{r^2-2Mr+a^2+Q^2}{r^2+a^2}\, du {-} 2\, dr\biggr] du~.
\label{metric2}
\end{eqnarray}
Furthermore, we assume that
$M$, $a$ and $Q$ are proportional to each other and 
linear functions of $u$, without loss of generality: 
\begin{eqnarray}
M(u) &=& M_0 ~(\mbox{constant}),\qquad u<u_0 \label{mass0} \\
&=& \mu u +b 
\equiv \tilde u, 
\qquad\quad~\; u_0< u \leq -b/\mu, ~\mu <0  \label{mass1} \\
&=& 0,\qquad\qquad\qquad~~ -b/\mu < u\\
a(u) &=& \lambda_1 M(u), \qquad\qquad~~~ 0 \leq \lambda_1 \leq 1\\
Q(u) &=& \lambda_2 M(u), \qquad\qquad~~~ 0 \leq \lambda_2 \leq 1 . \label{mass2}
\end{eqnarray}
That is, the black hole starts to evaporate at the retarded time $u=u_0$, and its mass
decreases at the rate $\mu$. 
{Although the angular momentum was held fixed in \cite{jing}, we make it time-dependent here as the derivation of the metric continues to hold with that generalization and furthermore,
the condition $\lambda_1, \lambda_2 \leq 1$ ensures that there are no naked singularities. }
Continuity of the functions $M(u),a(u),Q(u)$ Eqs.(\ref{mass0}-\ref{mass2}) guarantees that the various patches of the 
Penrose diagram that will smoothly join to each other \cite{pw}. 
We will comment on more general (non-linear) functions later. 

Next, to arrive at a set of coordinates which are smooth across the horizon, 
we adopt the procedure of \cite{pw}.
We first write metric (\ref{metric2}) in `double null' coordinates as:
\begin{eqnarray}
ds^2 = - \frac{g(\tilde u,r)}{\mu} d\tilde u dv ~,
\end{eqnarray}
where, 
\begin{eqnarray}
dv \equiv \frac{1}{g(\tilde u, r)} \left[ \left( \frac{r^2 - 2Mr +
(\lambda_1^2 + \lambda_2^2)M^2}{r^2 + {\lambda_1}^2 M^2}  \right) \frac{d\tilde u}{{\mu}} {-} 2dr \right]~.
\label{metric3}
\end{eqnarray}
To integrate the above, we first 
note that due to the linearity of functions in Eqs.(\ref{mass1}-\ref{mass2}), 
the functions multiplying $d\tilde u$ and $dr$ inside 
the square brackets in Eq.(\ref{metric3}) are homogeneous in $(\tilde u,r)$. 
Therefore the integrating factor $g(\tilde u,r)$ is given by:
\begin{eqnarray}
g(\tilde u,r) =  \left( \frac{r^2 - 2Mr +
(\lambda_1^2 + \lambda_2^2)M^2}{r^2 + {\lambda_1}^2 M^2}  \right) \frac{\tilde u}{{\mu}} {-} 2r~.
\end{eqnarray}
Then  
\begin{eqnarray}
\frac{\partial v}{\partial r} = \frac{r^2 + \lambda_1^2 \tilde u^2}{{-}r[r^2 + \lambda_1^2 \tilde u^2]
+ \frac{\tilde u}{2{\mu}} \left[ r^2 - 2\tilde u r + (\lambda_1^2 + \lambda_2^2)\tilde u^2 \right] } \label{v1} \\
\frac{\partial v}{\partial \tilde u} =
\frac{ \frac{1}{2a} ( r^2 - 2Mr + (\lambda_1^2 + \lambda_2^2)\tilde u^2) }{{-}r[r^2 + \lambda_1^2 \tilde u^2]
+ \frac{\tilde u}{2{\mu}} \left[ r^2 - 2\tilde u r + (\lambda_1^2 + \lambda_2^2)\tilde u^2 \right] }~, \label{v2}
\end{eqnarray}
whose denominators are cubic in $r$. Therefore the denominators have three roots $r_1,r_2,r_3$,
(with $r_1 >r_2 >r_3$) where $v$ has coordinate singularities, and 
\begin{eqnarray}
v &=& \sum_{i=1}^{3} A_i \ln(r-r_i)~.
\end{eqnarray}
Then the complete set of singularity free coordinates for the three patches can be
written as: 
{
\begin{eqnarray}
V_2(v) &=& e^{v/A_1} (r-r_1) (r-r_2)^{A_2/A_1}(r-r_3)^{A_3/A_1}~,\qquad r_2 \leq r < \infty  \\
V_1(v) &=& k_2 + (r_1-r)^{A_1/A_2} (r_2-r) (r-r_3)^{A_3/A_2}~,\qquad r_3 \leq  r < r_2 \\
V(v) &=& k_1 +(r_1-r)^{A_1/A_3} (r_2-r)^{A_2/A_3} (r-r_3)~,\qquad 0 \leq r < r_3 
\end{eqnarray}
}
where
{
\begin{eqnarray}
A_1 &=& \frac{r_1^2}{(r_1-r^2)(r_1-r_3)} + \lambda_1 \tilde u^2 ,~\qquad 
A_2 = \frac{-r_2^2}{(r_1-r^2)(r_2-r_3)} + \lambda_1 \tilde u^2 \\
A_3 &=& \frac{r_3^2}{(r_1-r_3)(r_2-r_3)} + \lambda_1 \tilde u^2 ~.
\end{eqnarray}
}
The constants $k_1$ and $k_2$ are determined by matching $V_2$ with $V_1$ in
$r_2 < r < r_1$ and $V_1$ with $V$ in $r_3 < r < r_2$ respectively.
If the denominator of Eq.(\ref{v1}) has one zero, say at $r_1$, such that it can be
factored as $(r-r_1)(r^2 + \beta r +\gamma)$, then
$v$ has one coordinate singularity, also at $r_1$. In this case, 
$v$ and $V$ are given by:
{
\begin{eqnarray}
v &=& A_1 \ln(r-r_1) + \frac{A_2}{2}\ln(r^2 + \beta r + \gamma) \nonumber \\
&+& \frac{2A_3-A_2\beta}{\sqrt{4\gamma - \beta^2}} \arctan \left( \frac{2r + \beta }{\sqrt{4\gamma-\beta^2}} \right) ~, 
\label{v4} \\
V(v) &=& e^{v/A_1}= (r-r_1)(r^2 + \beta r + \gamma)^{A_2/2A_1} \nonumber \\
&& \times \exp\left[ \frac{2A_3- A_2\beta }{A_1\sqrt{4\gamma-\beta^2}} \arctan 
\left( \frac{2r + \beta}{\sqrt{4\gamma-\beta^2}}\right) \right]~, \label{v5}
\end{eqnarray}
}
with
{
%
\begin{eqnarray}
A_1 &=& \frac{r_1^2 + \lambda_1^2 \tilde u^2}{r_1^2 + \beta r_1 +\gamma},~\qquad
A_2 = \frac{\beta r_1 + \gamma - \lambda_1^2 \tilde u^2}{r_1^2 + \beta r_1 +\gamma} \\
A_3 &=& \frac{r_1(-\gamma + \lambda_1 \tilde u^2)}{r_1^2 + \beta r_1 + \gamma}
\end{eqnarray}
}
%
{In either case, 
$r=0$ and its immediate neighbourhood is 
{\it timelike}}. 
For example, in the case of
Eqs.(\ref{v4}-\ref{v5}) above,
\begin{eqnarray}
ds^2 &=& - \frac{g(\tilde u,r) A}{V(\tilde u,r) {\mu} }~d\tilde u\, dV  \\
&\stackrel{\smash{r\rightarrow 0}}{\longrightarrow}& -\left( \frac{\lambda_1^2 + \lambda_2^2}{\lambda_1^2}\right) du^2 <0 ~.
\label{ds2}
\end{eqnarray}
{Note that the `ring singularity' of the metric is
at $r=0$ {\it and} $\theta=\pi/2$. Therefore, except for that singular point, the region at and near $r=0$ is regular and in fact one of low curvatures, except near the very end of the evaporation process.}

Next, we show that the apparent horizon is timelike as well, following the 
procedure adopted in \cite{kami}. 
We define the function:
\begin{eqnarray}
f(u,r)= r - M \mp (M^2 - Q^2 - a^2)^{1/2} ~,
\end{eqnarray}
such that the apparent horizons are at
\begin{eqnarray}
f(u,r)=0, ~r = M \pm (M^2 - Q^2 - a^2)^{1/2}~.
\end{eqnarray}
Now since $M$, $Q$ and $a$ are functions of $u$ 
the normal vector $n_a$ to the surface $f(u,r)=0$ {with components} given by
\begin{eqnarray}
n_u = f_{,u} 
= {-} \mu {\mp} \frac{\mu~( M - \lambda_1 a - \lambda_2 Q )}{ (M^2 - Q^2 - a^2)^{1/2} } ~,\qquad n_r= f_{,r}=1~,
\end{eqnarray}
{and} norm

\def \oldversion{ 
\begin{eqnarray}
n^2 = g^{ab}n_a n_b = 2 [ \mu \pm (\mu^2 -\alpha^2 -\beta^2)^{1/2}  ] >0~,
\label{n2}
\end{eqnarray}
%
{where $\alpha = \lambda_1\,\mu$ and $\beta = \lambda_2\,\mu$.}
Therefore, $n_a$ is spacelike and the apparent horizon is timelike.
}

\begin{eqnarray}
n^2 = g^{ab}n_a n_b = 2 { \mu [-1 \mp (1 -\lambda_1^2 -\lambda_2^2)^{1/2} } ] >0~,
\label{n2}
\end{eqnarray}
%
{where {we have used $\mu<0$ from Eq.~(\ref{mass1})}.}
Therefore, $n_a$ is spacelike and the apparent horizon is timelike.

With results Eqs.(\ref{ds2}), (\ref{n2})
and Eq.(\ref{evaphor}) in Appendix A, which shows that a slowly evaporating apparent horizon is also timelike, 
we can draw the Penrose diagram for the evaporating charged, rotating black hole (Figure 3). 
As one can see, there is a channel (outgoing null rays in red) which can carry classical as well as quantum information 
all the way from the 
interior to asymptopia, crossing the timelike horizon along the way. 

Several points are in order: \\
1. The generalization of our results to higher dimensions is straightforward. \\
2. Generalization for non-linear decay is also straightforward. 
One just needs to break-down the non-linear decay function, e.g.
$M(u)$ into a series of linear functions of the form
$M(u) = \sum_{i=1}^N (\mu_i u + b_i) 
\left[ \Theta( u_i+ \frac{1}{2}) - \Theta( u_i+ \frac{1}{2}) \right]$
($u_0 \leq u_i 
$), and similarly for
$a(u)$ and $Q(u)$, for large $N$. 
Now the apparent horizon
is piecewise continuous, and the previous conclusion continues to hold. 
\\
3. The condition $\theta =0$ in Eq.(\ref{metric2})
is not restrictive, since our aim was to show that there exists at least one such outgoing information channel. 
{Furthermore, there is no limit in principle to the amount of information that can exit through this channel.}
\\
4. Our work sheds no light on the resolution of singularities.
This is still to be addressed by 
quantum gravity theories \cite{st,lqg}, or via other approaches \cite{qre}. Note however that the information escape channel that we described is one of low curvature and hence our conclusions are robust. Our calculations also suggest that the singularity is always covered by the apparent horizon, until the end of the evaporation process. 
\\
5. Since proposals such as the fuzzball and firewall deal purely 
with quantum information, and in the context of Hawking radiation,
there is no contradiction between those
and our results here \cite{fuzz,fire}. On the contrary, our approach complements 
these others. 
\\
%
{6. It is known that matter flowing into the dynamic charged and rotating black hole may cause instability due to mass-inflation, resulting in the nature of the singularity changing \cite{israel,rossi}. However, we do not anticipate such a problem in our case since (a) our results pertain to the region of low curvature (not singularity),
and (b) we have outgoing as opposed to infalling matter. That said a careful study of the effect for the spacetime under consideration is warranted and we hope to report on this in the future. }
\\
{7. If the evaporation process stops for any reason before the black hole evaporates completely, our Fig.3 would change and revert back to Fig.2. However, since the Hawking temperature remains non-zero throughout, we do not consider this possibility.} \\
8. It will be interesting to compute the correlations between outgoing Hawking radiation and infalling particles. We leave it to a future publication.

{To summarize, we have shown here that 
practically for all black holes in nature, there exists a channel through which classical or quantum information can evaporate.
%
%
Although our work does not
completely resolve the problem of information loss, it complements
other approaches in opening up a new information recovery channel.
More work needs to be done to determine the extent of information that can be extracted via this channel, and its interplay with the 
quantum resolution of the singularity. 
We hope to report on these and related issues elsewhere.
}

\begin{figure}[htp] 
\centering
{\includegraphics[width=0.35\textwidth]
{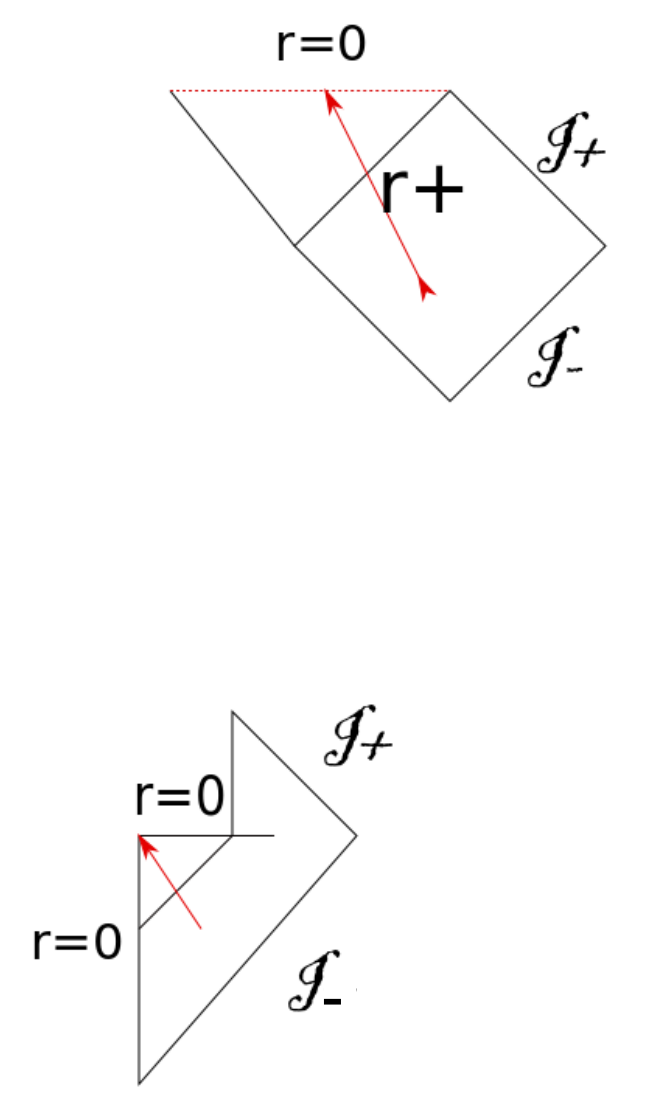}}
\caption{Schwarzschild Penrose diagram}
\end{figure}

\begin{figure}[htp] 
\centering
{\includegraphics[width=0.4\textwidth]
{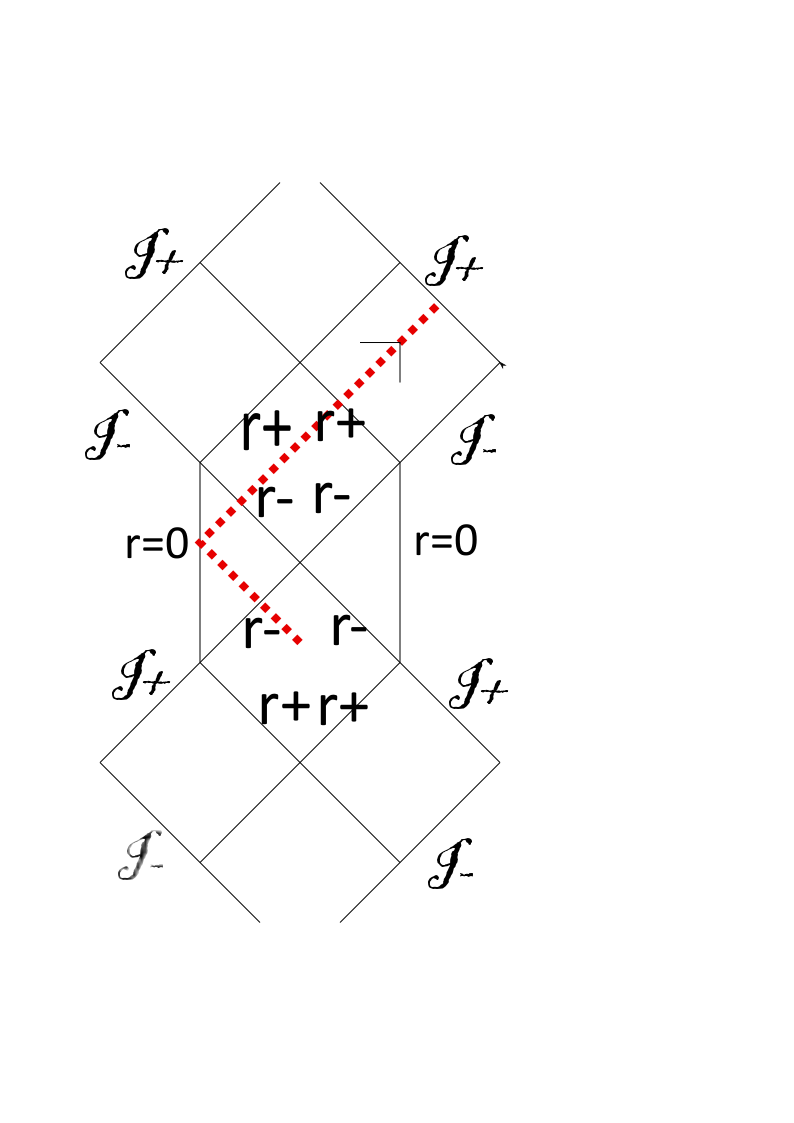}}
\caption{Kerr Penrose diagram}
\end{figure}

\begin{figure}[htp] 
\centering
{\includegraphics[width=0.4\textwidth]
{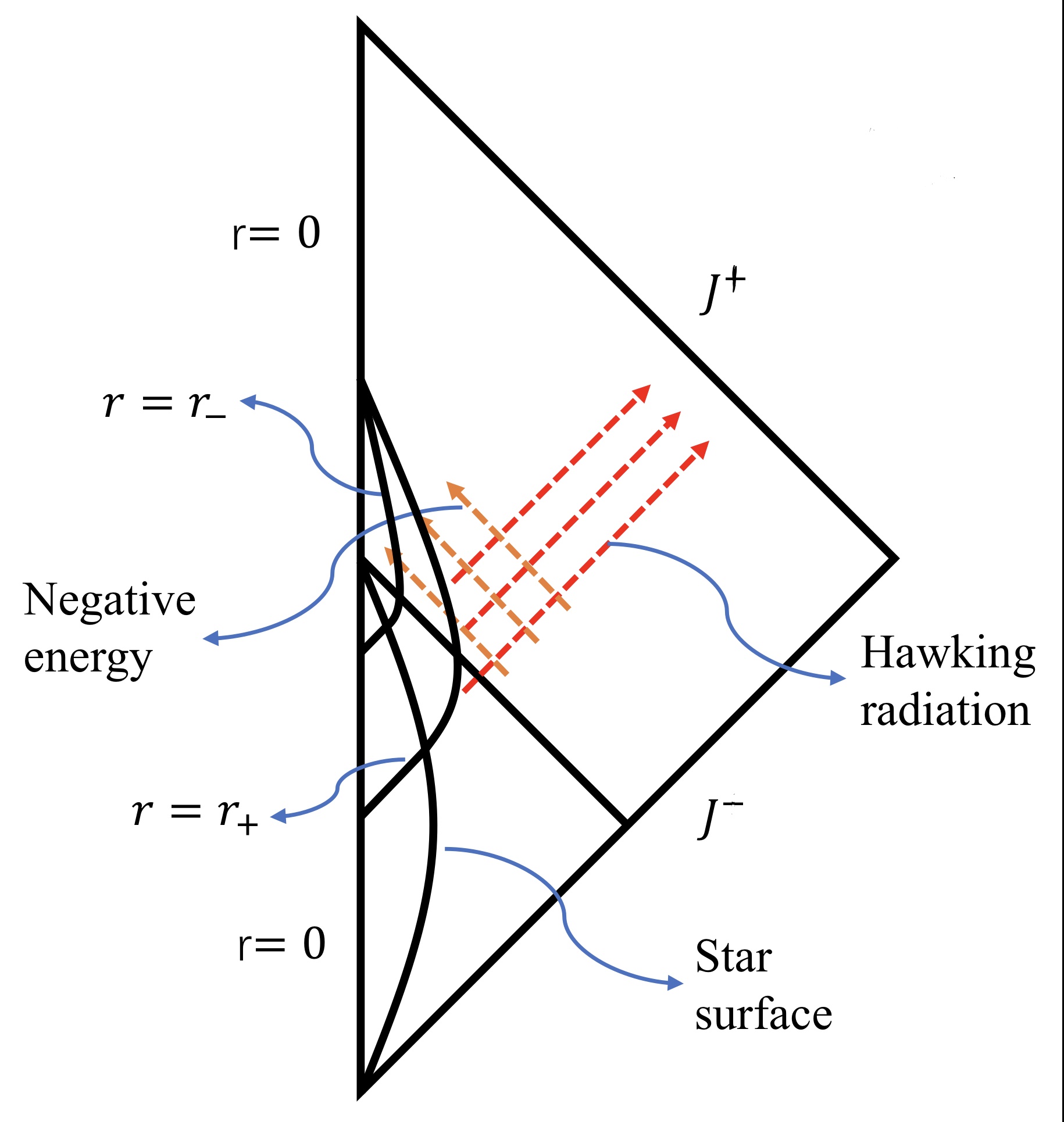}
}
\caption{Evaporating black hole 
Penrose diagram}
\end{figure}

\subsection{Acknowledgement}
%
\vspace{-2ex}
\par\noindent
This work was supported by the Natural Sciences and Engineering Research Council of Canada.
We thank the anonymous referee for useful comments which have helped improve the paper. 
%


\section*{{Appendix}}
\label{AppendixA}

\subsection*{{Appendix A: The metric in our manuscript represents a black hole}}

{
To show that our metric represents an evaporating black hole, we calculate the expansion of the null normal vectors for the rotating charged Vaidya-type black hole, see Eq.~(1) in the manuscript.  
}

{
Suggested by the rotating charged stationary black hole \cite{Newman}, we may conjecture that the outward null vector $l^\mu$ and the inward null vector $n^\mu$ for the rotating charged Vaidya-type black hole are
}
\begin{equation}
l^\mu=\Bigl( \frac{r^2 +a^2}{\sigma^2},\frac{\triangle}{2 \sigma^2},0,\frac{a}{\sigma^2} \Bigr) \;\;\;\; \text{and} \;\;\;\; n^\mu=(0,-1,0,0) .
\end{equation} 
where $\sigma^2=r^2 +a^2 \cos^2 \theta$ and $\triangle =r^2 +a^2-2Mr+Q^2$.
It is easy to check that

\begin{equation}
l^\mu l_\mu =0 , \;\;\; n^\mu n_\mu=0 \;\;\; \text{and} \;\;\; l^\mu n_\mu=-1
\end{equation}
for the Vaidya-type metric in our manuscript. Then we calculate the expansion for both $l^\mu$ and $n^\mu$ by $\theta^{(A)} = A_{\mu ; \nu} \sigma^{\mu\nu}$, where $\sigma^{\mu\nu} = g^{\mu\nu} + l^\mu n^\nu + n^\mu l^\nu$. Finally, we obtain

\begin{equation}
\theta^{(l)} = \frac{ r \triangle + 2 \, a \,a'(u) (r^2 + a^2 \cos ^2(\theta ))}{(r^2 + a^2 \cos ^2(\theta ))^2}  \;\;\;\; \text{and} \;\;\;\;  \theta^{(n)} = -\frac{2 r}{r^2 + a^2 \cos ^2(\theta )} .
\label{expansion}
\end{equation}

%
{
From Eq.~(\ref{expansion}) we see that $\theta^{(l)}>0$ and $\theta^{(n)}<0$ when $r$ is larger than the apparent horizon and $\theta^{(l)}=0$ defines the apparent horizon. Thus, the Vaidya-type metric in our manuscript represents a black hole.
}

{
The apparent horizon for a stationary rotating charged black hole is defined by $\triangle=0$. Here, our calculations show that the time dependence of the angular momentum $a(u)$ makes the apparent horizon shift slightly from the stationary case with the size of the shift depending on the rate of change of $a(u)$ with respect to $u$.
}

{
If we insert our linear functions for $M$, $Q$ and $a$, Eqs.~(3-6) in the manuscript, into the expansion of the outward null normal vector and suppose the rate of change is very slow ($\mu$ is very small), we can obtain the corrected apparent horizon for $\theta=0$ as
}
\begin{equation}
R_\pm = r_\pm + \frac{a \lambda_1 (r_\pm^2+a^2)}{r_\pm^2-(M+2a\mu \lambda_1)r_\pm} \mu + O(\mu^2) .
\label{evaphor}
\end{equation}
Since we have proved $r_\pm$ is timelike, the apparent horizon of a slowly evaporating rotating charged black hole should also be timelike.

\subsection*{
{Appendix B: The metric in
\cite{Parikh1999}
represents a white hole}}

{ As far as we know, the first one try to do a similar analysis as us, though limited to non-rotating scenarios, are Parikh and Wilczek \cite{Parikh1999}. However, although they claimed that they analyzed the case of an evaporating charged black hole, the metric used in their paper is in fact the metric of a charged white hole. They study the metric \cite{Parikh1999}}
\begin{eqnarray}
ds^2 = -\biggl(1-\frac{2Mr-Q^2}{r^2}\biggr)du^2 - 2du\, dr + r^2 ( d\theta^2 + \sin^2\theta\, d\phi^2), 
\label{Parikh}
\end{eqnarray}
where $M=M(u)$ and $Q=Q(u)$ denote smooth decreasing functions of the retarded time $u$. It is easy to see that Eq.~(\ref{Parikh}) represents a white hole by taking $Q=0$ \cite{Poisson2004}. To exactly show that Eq.~(\ref{Parikh}) is the metric of a white whole, we now calculate the expansion of the null normal vector of this metric.

{
Since this metric is spherically symmetric, it is easy to figure out that the outward null normal vector $l^\mu$ and the inward null normal vector $n^\mu$ for this metric are
}
\begin{equation}
l^\mu=(0,1,0,0)  \;\;\;\; \text{and} \;\;\;\; n^\mu= \Bigl( 1,-\frac{r^2 -2Mr+Q^2}{2 r^2},0,0 \Bigr).
\end{equation} 
{
It is then easy to check that
}
\begin{equation}
l^\mu l_\mu =0 , \;\;\; n^\mu n_\mu=0 \;\;\; \text{and} \;\;\; l^\mu n_\mu=-1 .
\end{equation}
{
Finally, we calculate the expansion for both $l^\mu$ and $n^\mu$ by $\theta^{(A)} = A_{\mu ; \nu} \sigma^{\mu\nu}$, where $\sigma^{\mu\nu} = g^{\mu\nu} + l^\mu n^\nu + n^\mu l^\nu$, to obtain
}
\begin{equation}
\theta^{(l)} = \frac{2}{r} \;\;\;\; \text{and}   \;\;\;\;  \theta^{(n)} = - \frac{ r^2 -2Mr+Q^2 }{r^3} .
\label{expansionP}
\end{equation}

{
From Eq.~(\ref{expansionP}) we know that $\theta^{(l)}>0$ and $\theta^{(n)}<0$ when $r$ is larger than the apparent horizon. However, it is the vanish of $\theta^{(n)}$ defines an apparent horizon. Thus, the Vaidya-type metric in Parikh and Wilczek's paper represents a white hole.
}

\end{document}